\documentclass[
11pt,%
tightenlines,%
twoside,%
onecolumn,%
nofloats,%
nobibnotes,%
nofootinbib,%
superscriptaddress,%
noshowpacs,%
centertags]%
{revtex4-1}
\usepackage{ljm}
\usepackage{diagbox}
\usepackage{listing}

\begin{document}
\titlerunning{Efficient reduction of Feynman integrals on supercomputers} 
\authorrunning{A.V. Belitsky, A.A. Kokosinskaya, A.V. Smirnov, V.V. Voevodin, M. Zeng} 

\title{Efficient reduction of Feynman integrals on supercomputers}

\author{\firstname{Andrei V.}~\surname{Belitsky}}
\email[E-mail: ]{andrei.belitsky@asu.edu}
\affiliation{Department of Physics, Arizona State University, Tempe, AZ 85287-1504, USA}

\author{\firstname{Anna A.}~\surname{Kokosinskaya}}
\email[E-mail: ]{legko.zapomnit@mail.ru}
\affiliation{Moscow State University, Research Computing Center, 119992 Moscow, Russia}

\author{\firstname{Alexander~V.}~\surname{Smirnov}}
\email[E-mail: ]{asmirnov80@gmail.com}
\affiliation{Moscow State University, Research Computing Center, 119992 Moscow, Russia}
\affiliation{Moscow Center for Fundamental and Applied Mathematics, 119992 Moscow, Russia}

\author{\firstname{Vadim V.}~\surname{Voevodin}}
\email[E-mail: ]{vadim_voevodin@mail.ru}
\affiliation{Moscow State University, Research Computing Center, 119992 Moscow, Russia}

\author{\firstname{Mao}~\surname{Zeng}}
\email[E-mail: ]{zengmao@hotmail.co.uk}
\affiliation{Higgs Centre for Theoretical Physics, University of Edinburgh, Edinburgh, EH9 3FD, UK}


\firstcollaboration{(Submitted by Alexander Smirnov) } 

\received{} 

\begin{abstract} 
Feynman integral reduction by means of integration-by-parts identities is a major power gadget in a theorist toolbox indispensable
for calculation of multiloop quantum effects relevant for particle phenomenology and formal theory alike.  An algorithmic approach consists of solving a large sparse non-square system of homogeneous linear equations with polynomial coefficients. While an
analytical way of doing this is legitimate and was pursued for decades, it undoubtedly has its limitations when applied in complicated circumstances. 
Thus, a complementary framework based on modular arithmetic becomes critical on the way to conquer the current `what is possible' frontier.
This calls for use of supercomputers to address the reduction problem. In order to properly utilize these computational resources, one has to 
efficiently optimize the technique for this purpose. Presently, we discuss and implement various methods which allow us to 
significantly improve performance of Feynman integral reduction within the FIRE environment.
\end{abstract}

\maketitle

\section{Introduction}

Successful verification of particle physics models demands accuracy of theoretical predictions on par with experimental measurements. This 
calls for calculations of high-order quantum effects in physical observables. These are readily encoded with Feynman Integrals (FIs)
\begin{align}
G^{(L)}_{a_1, \dots, a_N} (p_1, \dots, p_E) = \int \frac{d^D k_1}{i \pi^{D/2}} \dots \frac{d^D k_L}{i \pi^{D/2}} D_1^{- a_1} \dots D_N^{- a_N}
\end{align}
in a $D = 4 - 2 \varepsilon$ dimensionally regularized theory, where the number $L$ of $D$-fold integrations is determined by the 
perturbative order in coupling constants. The integrand is built out by a product Lorentz-invariant bilinear functions $D_i$ of loop $k_i
=q_i$ ($i = 1, \dots, L$) and external $p_i = q_{i + L}$ ($i = 1, \dots, E$) particles' momenta raised to integer powers $a_i \in \mathbb{Z}$. 
For a given kinematical configuration, the complete basis of $D_i$ functions is spanned by $N = L (L+1)/2 + LE$ elements and the 
number of FIs thus grows as $\mathbb{Z}^N$. 

To date, the most successful tool to find a minimal set, if it exists, of FIs for a given graph, is based on the integration-by-parts (IBP) identites 
\cite{Chetyrkin:1981qh},
\begin{align}
\label{IBPs}
\int \frac{d^D k_1}{i \pi^{D/2}} \dots \frac{d^D k_L}{i \pi^{D/2}} \partial_i \cdot q_j D_1^{- a_1} \dots D_N^{- a_N} = 0
\quad \mbox{for}\quad
i=1, \dots,L\, ; \ j=1,\dots, L+E
\, ,
\end{align}
which possess vanishing right-hand side only away from the four-dimensional space-time, i.e., $\varepsilon \neq 0$. This is the first reason to 
focus on $D\neq 4$. The second reason being that massless gauge theories develop infrared and (generally) ultraviolet divergences which 
require a regulator to make FIs well defined. As it is obvious from the representation (\ref{IBPs}), for a given point in the $\mathbb{Z}^N$ space 
of FIs there are $L(L+E)$ IBPs. The derivatives in the integrand shift the $a_i$ indices by $\sigma_i = \pm 1, 0$ and yield an underdetermined 
system of homogeneous linear equations for FIs. This system, once solved, allows one to deduce a minimal set of undetermined FIs 
known as the Master Integrals (MIs). A priori, it is completely unclear whether that number is even finite given there are infinitely many relations for 
infinitely many integrals as the integers $a_i$ are not restricted. However, this was affirmatively demonstrated in Ref.\ \cite{Smirnov:2010hn} 
using algebro-geometric techniques. Unfortunately, the proof is non-constructive and cannot be used either to help in solving the reduction 
problem or even to determine the number of MIs for a given family of FIs. 

The procedure is thus to efficiently solve a huge sparse non-rectangular homogeneous system of linear equations 
\begin{align}
\label{IBPs2}
\sum_{\{\sigma_1, \dots, \sigma_N\}} c_{\sigma_1, \dots, \sigma_N} G^{(L)}_{a_1 + \sigma_1, \dots, a_N  + \sigma_N} = 0
\end{align}
with polynomial coefficients $c_{\sigma_1, \dots, \sigma_N}$, being functions of the space-time dimension $D$ and  kinematical 
invariants $s_{ij} \equiv (p_i + p_j)^2$. This is a classical problem in linear algebra and belongs to centuries-old books. However, a practical tool 
applicable to field-theoretical setups was suggested relatively recently starting with the work by Laporta \cite{Laporta:2000dsw}. It is based on a 
version of the well-known Gaussian elimination technique and takes into account a chosen priority of points in the  $\mathbb{Z}^N$ space of FIs. 
It roughly consists in starting with a sector containing the smallest allowed number of $D_i$'s, solving for a subset of FIs and subsequently 
substituting these into FIs of the next level in complexity containing more $D_i$'s etc.\ etc. This is done analytically in $D$ and $s_{ij}$ 
without specifying their numerical values and relies on a heavy computer use. This technique is currently implemented in quite a number 
of public and private computer codes such as, but not limited to, {\tt AIR} \cite{Anastasiou:2004vj}, {\tt FIRE} \cite{Smirnov:2023yhb}, 
{\tt LiteRed} \cite{Lee:2012cn} and {\tt Kira} \cite{Klappert:2020nbg}.

While the traditional IBP approach described in the previous paragraph was proven to be extremely successful in practical solution of problems
in high energy physics, in many circumstances it commands the use of computers with multiple CPUs and, what is worse, terabytes of RAM.
Nowadays, multiple CPUs is an industry standard for personal computers and work stations, however, the need for high-memory nodes is a 
user-unfriendly constraint, either in their very availability even on superclusters or long queue waiting times for their access. And even if both of these 
conditions are fulfilled, an IBP reduction for a complex Feynman graph could take months of computer time without any warranty for its 
success in the end. The reason for this is clear from our brief description of the Laporta algorithm alluded to above. Namely, at every step of the 
system-solving procedure, one needs to call for an external simplification library of rational fractions, which are generated in coefficients 
accompanying MIs, and bring the fraction $c/c'$ to a canonical form by getting rid of common factors in its numerator and denominator. This is a 
time-consuming process and turns out to be the bottleneck for the entire endeavor. A recent work \cite{Mokrov:2023vva} introduced efficient 
libraries to partially alleviate the problem. However, one definitely needs to look for alternative approaches which would bypass the need for 
excessive use of RAM required to handle swelling intermediate expressions.

A new framework is especially indespensable for physical observables depending on multiple scales such as high-multiplicity scattering amplitudes 
and form factors, see, e.g., recent \cite{He:2022ctv,Belitsky:2023gba}. These involve multiple $s_{ij}$-invariants and, if one needs to consider 
off-shell kinematics, depend in addition on external particles' virtualities $p_i^2 \neq 0$. The presence of the latter makes the resulting 
IBP reduction practically insurmountable for direct analytical calculations. In fact, a technique which overcomes this predicament is known for almost 
a decade now
\cite{vonManteuffel:2014ixa,Peraro:2016wsq,Peraro:2019svx,Klappert:2019emp,Klappert:2020aqs,Laurentis:2019bjh,DeLaurentis:2022otd,Magerya:2022hvj,Belitsky:2023qho} and is based on modular arithmetic. Its basic idea is quite simple: substitute numerical values for all variables involved
and then solve the resulting system of IBP equations (\ref{IBPs2}). This problem is naively much easier since the emerging coefficients are now 
numbers rather than ratios of lengthy polynomials. Still working with rational numbers may not be efficient enough for after the number coefficients 
can swell uncontrollably as well in the course of the Gaussian elimination and, therefore, not fit into the machine arithmetic format, either 64-bit or even 128-bit. 
To practically work with these numbers, one would need special libraries using long-integer arithmetic. This would make the reduction inefficient, 
defeating the purpose from the get-go. This impels one to switch to the modular arithmetic on the finite field $\mathbb{Z}_p$ by choosing the prime $p$ not exceeding $2^{64}$ instead of working in the field of rationals $\mathbb{Q}$. This is a standard practice in computer science as was reminded to the physics community in Ref.\ \cite{Kauers:2008zz}. In this set-up, all the expansion coefficients 
$c_{\sigma_1, \dots, \sigma_N}$ fit into the machine arithmetic and, as a consequence, an IBP reduction can be effectively run very fast.

A single point or a few (modular or not) in the $\mbox{\boldmath$v$}= (D,s_{ij})$ parameter space will not suffice to get the full parametric dependence of $c$'s. The question 
is then how many does one need to unambiguously reconstruct the polynomials with rational number coefficients without any prior knowledge of their order or the number of nonvanishing  terms, and, in addition, accompanying expansion coefficients restored in turn 
from finite fields when these are used. This last step relies on classical ideas based on the well-known Chinese Remainder Theorem  to combine short primes, which fit into a processor word, into large numbers making subsequent rational reconstruction from $\mathbb{Z}_{p_1 \times \dots \times p_n}$ by the Extended Euclidean Algorithm unambiguous \cite{Gathen:2013}. 
The reconstruction from rational numbers to rational functions is far more complex and was addressed in Refs.\
\cite{vonManteuffel:2014ixa,Peraro:2016wsq,Peraro:2019svx,Klappert:2019emp,Klappert:2020aqs,Laurentis:2019bjh,DeLaurentis:2022otd,Magerya:2022hvj,Belitsky:2023qho} making use of a variety of methods. Some of these are more time consuming than the others. Thus the goal of an efficient reconstruction approach is then to require (i) the 
least possible number of blackbox samples (BBS) and (ii) a feasible restoration of mutivariate polynomials from these samples. Obviously, these two 
steps combined have to take significantly less cumulative time than the actual analytical reduction (if it is at all possible). 

A very attractive feature of the modular approach is its amenability to parallelization: IBP reductions for various values of 
parameters over the finite field $\mathbb{Z}_{p}$ can be performed independently from one another. And while one might need thousands or even millions of 
reduction results, this approach ideally fits for supercomputer use, if properly organized. The IBP program {\tt FIRE}, developed in Ref.\ 
\cite{Smirnov:2008iw}, already had the modular approach built-in in the code starting from the version {\tt 6.0} \cite{Smirnov:2019qkx}. However, 
its previous implementation was not sufficiently practical, in spite of being successfully used once in Ref.\ \cite{Lee:2019zop}, due to unbefitting 
reconstruction methods. This last problem was resolved in Ref.\ \cite{Belitsky:2023qho} with the development of a novel technique for dense interpolation dubbed the balanced 
reconstruction, however, it was not coded or optimized for supercomputers. This drawback will be overcome in this paper. As a benchmark, we will use 
an example from a recent study of non-planar FIs contributing to a three-leg off-shell form factor \cite{Belitsky:2023gba}, in particular, one of the 
most time-consuming IBP reductions of 38 level-7 FIs, arising in differential equations, down to the level 3 MI $G^{(2)}_{0, 1, 0, 1, 1, 0, 0, 0, 0}$.
The standard analytical route was possible there but it was taking more than ten days and required extremely high memory use. Below we will report 
on the performance upgrade and improvements of the current version of {\tt FIRE} which yielded a significant overall time reduction by almost an 
order of magnitude compared to the analytical approach.

Below to provide an in-depth analysis of various optimization techniques used to make the reconstruction of IBP identities more efficient, including the use of coding in C++, memory management and use of external libraries like Flint.

\section{Reconstruction in c++}

Up to now, all public versions of {\tt FIRE} capable of reconstruction, the initial \cite{Smirnov:2019qkx} and the more recent \cite{Belitsky:2023qho}, 
relied on codes written in {\tt Wolfram Mathematica}. The latter is a commercial software which is typically not installed on supercomputers.
Thus, in Ref.\ \cite{Lee:2019zop} alluded to above, proliferating back-and-forth transfers had to be done between a supercomputer, used to generate 
sampling points, and a local machine, used to reconstruct from these, creating a logistical burden. While it was borderline feasible for two-variable 
observables studied there, it would be highly impractical for more. In the current version of {\tt FIRE}, used throughout this work, the balanced 
reconstruction \cite{Belitsky:2023qho} is coded in C++. A typical line in a script, which provides a brief glossary of options, reads\footnote{Depending 
on the {\tt Slurm} scheduler, {\tt mpirun} can be superseded by 
{\tt srun} and the accompanying flag for the number of cores {\tt NC0} used changes 
from {\tt -np} to {\tt -n}.
}
{\footnotesize
\begin{verbatim}
    mpirun -np NC0 FIRE6_MPI --calc ${CALC0} --reconstruct --variables v1_v2_v30 --newton N1_N2 
                             -c config "v10:v1T;v20:v2T;v30:v30;P0"
\end{verbatim}
}
This calls for the generation of BBS with {\tt FIRE} for the $\mbox{\boldmath$v$}$-variables $v_1$, $v_2$ and $v_3$ in their Thiele ranges $[v_{1,0},v_{1,\rm T}]$, 
$[v_{2,0},v_{2,\rm T}]$ (the corresponding Newton ranges of the balanced reconstruction are defined by the first $N_1$ and $N_2$ elements in these) and the last variable $v_3$
in this example is fixed at $v_{3,0}$. The number of primes to be used in the rational reconstruction is set by the value $P_0$.
Moreover, the code is organized in a manner that automatically shuts the production of sampling points down when their sufficient number is 
accumulated for a successful unambiguous reconstruction. This is activated by adding the flag \verb|--abort|. For large-scale computations, the number 
of generated files could potentially exceed the storage capacity of Linux file systems and {\tt FIRE} can be instructed to delete BBS no longer needed in 
the course of a computation with \verb|--delete_tables|. Last but not least, external computer algebra systems used in intermediate simplifications is 
selected with the option \verb|CALC0|.

\section{Optimization by changing the order of reconstructions}
\label{ReconOrderSect}

According to the general logic spelled out in the Introduction, and made explicit in its implementation in the sample script of the previous 
section, one starts the process by substituting integer values for all $\mbox{\boldmath$v$}$-variables in IBP identities and then proceeds with a choice of prime 
numbers for the modular arithmetic. So it only appears natural to perform the reconstruction in the opposite order: (let us dub it as the `Beast' mode\footnote{Ugly but works.}) primes $\to$ rational coefficients $\to$ rational functions. This approach was originally implemented in {\tt FIRE} with {\tt Mathematica} \cite{Belitsky:2023qho} and its private 
C++ realization. This route was rather straightforward since it did not require employment of modular-polynomial arithmetic being that the 
reconstruction of rational numbers from primes was done first. Then an external library was called for operations on polynomial and rational 
function.

However, the reader can already anticipate a drawback intrinsic to the above approach. Namely, when reconstructing a rational number from its 
projections over modular fields, the number of primes required for its unique restoration heavily depends on the `size' of the former, i.e., the maximum 
of absolute values of its numerator and denominator. The bigger the size is, the more modular values one needs and hence more IBP reductions to perform. 
So this quickly becomes an issue as 
the expansion coefficients $c$ in the IBPs (\ref{IBPs2}) are polynomial in the $\mbox{\boldmath$v$}$-variables with integer-valued expansion coefficients $C_{i_1 i_2, \dots}$, which are sought for,
\begin{align}
\label{cs}
c = \sum_{i_1, i_2, \dots} C_{i_1, i_2, \dots} v_1^{i_1} v_2^{i_2} \dots
\end{align}
and after the variable substitutions and summation, the result is in general much bigger in size than the initial values of $C_{i_1, i_2, \dots}$ that one 
started with. 

A panacea to the above predicament was found in changing the order of the two steps by first reconstructing rational functions with their 
$C$-coefficients in a modular field and then restoring them from primes to rationals, i.e., (let us call it the `Beauty' mode\footnote{Elegant and efficient.}) primes $\to$ rationals functions $\to$ 
rational coefficients. The
first step here is done with the very same balanced method of Ref.\ \cite{Belitsky:2023qho}, while in the second step, we parse the reconstructed 
rational functions over primes by analyzing their structure: We throw away exceptional cases where something was accidentally canceled out due 
to an unfortunate choice of variables and consider only rational functions possessing the same monomial structure both in their numerators and 
denominators but, obviously, different expansion coefficients over primes. Then for each set of these, accompanying a given monomial, we finally run 
the rational reconstruction.

Another feature of paramount importance which makes the Beauty preferable to the Beast mode is the fact that the number of prime values needed for an unambiguous
reconstruction of rationals is not only smaller but stable as well, i.e., it does not depend on numerical values of variables used because the 
expansion coefficients $C_{i_1, i_2, \dots}$ in the original function are constants to start with. This eliminates the guessing game from
the robust estimate for seeding of BBS. For the Beast mode it is almost next to impossible to predict the number of necessary reductions in advance, only an upper 
limit estimate, which yields quite an overestimation, is possible. The new way allows us to minimize the number of IBP reductions. This will be discussed in the following section. 

\begin{table}[t]
\begin{center}
\begin{tabular} { | l | l | l | l | l|}
\hline
\multicolumn{1}{|l|}{Method} &
\multicolumn{2}{l}{Beast mode} &
\multicolumn{2}{|l|}{Beauty mode} \\
\hline
\diagbox{vars}{BBS} & \# p's & \# r's  & \# p's & \# r's \\
\hline
\hline
$d$ & 11 & 13$_{\rm T}$  & 11 & 13$_{\rm T}$ \\
$w$ & 14 & 100$_{\rm T}$ & 11 & 100$_{\rm T}$ \\
$v$ & 15 & 106$_{\rm T}$ & 11 & 106$_{\rm T}$ \\
$u$ & 18 & 128$_{\rm T}$ & 11 & 128$_{\rm T}$ \\
($d$,$w$) & 15 & 13$_{\rm T}$$\times$51$_{\rm N}$ & 11 & 13$_{\rm T}$$\times$51$_{\rm N}$ \\
($d$,$w$,$v$) & 16 & 13$_{\rm T}$$\times$51$_{\rm N}$$\times$54$_{\rm N}$ & 8 & 13$_{\rm T}$$\times$51$_{\rm N}$$\times$54$_{\rm N}$ \\
($d$,$w$,$v$,$u$) & 19 & 13$_{\rm T}$$\times$51$_{\rm N}$$\times$54$_{\rm N}$$\times$65$_{\rm N}$ & 7 & 13$_{\rm T}$$\times$51$_{\rm N}$$\times$54$_{\rm N}$$\times$65$_{\rm N}$ \\
\hline
\end{tabular}
\end{center}
\caption{\label{CompOrderTable} Comparison of the Beast  (primes $\to$ rational coefficients $\to$ rationals functions) and Beauty modes (primes $\to$ rationals functions $\to$ 
rational coefficients) restorations. Labels on the number of rational numbers required for polynomial reconstruction stand for Thiele (T) and balanced Newton (N), respectively.}
\end{table}

Before closing this section, let us demonstrate the realization of these improvements with our benchmark. The problem involves five 
variables $\mbox{\boldmath$v$} = (D,w,v,u,m)$, i.e., the space-time dimension $D$, three Madelstam invariants $u,v,w$ and a virtuality $m$. Since FIs 
are homogeneous functions of the $(w,v,u,m)$ variables, one can safely set one of them equal to one, say $m=1$, and restore it at the very end from 
naive dimensional counting. To make a robust estimate for the number of sample points required in each of these, we ran trial IBP reductions (each 
takes about a minute or two on a small number of cores $\sim 50$), which are summarised in the first four rows of Table \ref{CompOrderTable}. We used 
there the following starting value for $\mbox{\boldmath$v$}_0 = (40,30,20,10,1)$ and relied on {\tt CALC0}$=${\tt Flint} \cite{Flint} as a simplifying 
external software. We immediately observe that while the number of necessary modular values in the finite field grows substantially with increasing 
initial values of variables for the Beast mode, it stays the same for the Beauty. The reason for this was already elucidated in the previous paragraph. 
As we enlarge the set of to-be-reconstructed variables, for the Beast mode we always have to employ the largest number of primes among those needed for 
individual restorations. Moreover, it was even necessary to add an extra one to warrant successful runs. On the contrary, in the Beauty mode, the number 
of primes decreases with the increasing number of variables subject to restoration. This trend is obvious from the right-hand side of Eq.\ (\ref{cs}) 
since, as we first perform the rational function reconstruction, the size of accompanying coefficients gets smaller\footnote{The $C$ coefficients are 
no longer multiplied by high powers of integers!}. This yields an e-fold, i.e., $19/7 \simeq 2.7$, reduction in the overall number of required sample
points. In practical applications, it is advisable to increase the Newton's range for all variables by a few values to ensure successful reconstruction. The reason being that sometimes tables may arrive corrupted, or not at all, from some nodes and, therefore, 
unsuitable for further use. With this caveat in mind, for the full ($d$,$w$,$v$,$u$)-variable computation, performed on 1024 cores of the ASU's 
{\tt Sol} supercomputer \cite{ASUSol}, we clocked the two reconstructions at 2-10:26 ($\sim 58$ hours) and 1-04:22 ($\sim 28$ hours), respectively. 
This is more than a factor of two speedup, making the Beauty mode copacetic. Further details will be provided in Section \ref{FlintSect} below.

\section{Analyzing and Optimizing application performance} 

The methods discussed in the previous sections were intended to minimize the number of reductions needed to produce the reconstruction, so they belong to algorithmic improvements. In this section, we would like to focus on another important area of improvement -- software optimizations based on application performance analysis.

The program analysis was performed on the Lomonosov-2 supercomputer. In most cases, the \texttt{pascal} partition was used, in which each node contains one 12-core Intel Xeon Gold 6126 CPU with 92 GB of RAM. The application ran on 210 processors, with Open MPI 1.8.4 being used. The average execution time of the analyzed application is 115 s.


First, an analysis of working with MPI was carried out. The program operates in a master-worker paradigm, where an MPI process with rank zero is assigned a role to distribute tasks and control the execution of other processes. Analysis of MPI usage in the application was performed using mpiP 3.5 profiler~\cite{mpip}.

The results showed that MPI operations occupy only a small portion (<1\%) of the overall execution time. It was observed that the master process spends more than 96\% of the time waiting for the completion of tasks by other processes. But if we start distributing the payload to the master process as well, this does not lead to noticeable redistribution load, but may cause the master process to respond more slowly to worker process requests. Moreover, such a small fraction of the time spent on MPI operations suggests that working with MPI is clearly not a bottleneck in this program.

\begin{figure}[!h]
\includegraphics[width=1.0\textwidth]{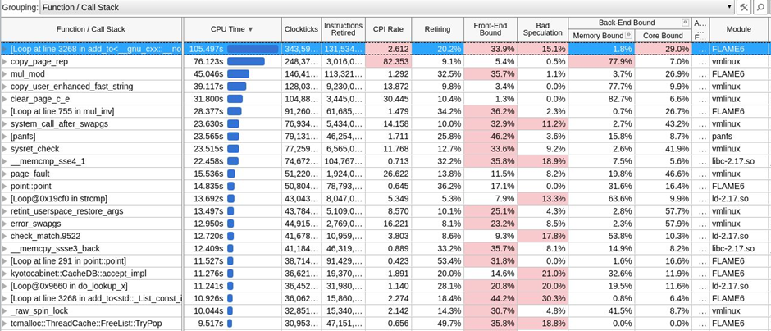}
\caption{The most computationally expensive program fragments and their performance metrics, analysis performed using Intel VTune Profiler}
\label{fig:vtune}
\end{figure}

Next, a general performance analysis of individual processes was carried out using the Intel VTune Profiler tool, version 2019.5. Fig.~\ref{fig:vtune} shows our analysis of the most computationally expensive fragments (functions, loops) used in the application. In each line, the leftmost column corresponds to the fragment name, the time for executing this fragment is indicated in ``CPU Time'' column, and on the right side the columns show the values of metrics from the Top-down approach~\cite{top-down}. A detailed description of Top-down metrics can be found in Intel documentation~\cite{top-down-metrics}. The rightmost column shows whether this fragment is executed in the program itself (the value ``FLAME6'') or whether it belongs to an external library that was called from the program.

Many of the most frequently used functions are Front-End Bound (corresponding cells are marked red), which means that the processor quite often does not manage to promptly preprocess instructions for their execution in functional units, leading to them being idle waiting. Next, one can notice that there are functions from the Linux kernel module \texttt{vmlinux} for paged memory organization (for example, \texttt{copy\_page\_rep}, \texttt{clear\_page\_c\_e}), which also have a high Memory Bound indicator, i.e., during their execution the processor is often idle waiting for data from the memory. In addition, many external functions have a high Bad Speculation value. This indicates that the processor is often busy executing instructions that then turned out to be unnecessary (the most common reason is incorrect branch prediction). From the other side, the operations \texttt{add\_to} (line 1), \texttt{mul\_mod} (line 3) and \texttt{mul\_inv} (line 6), which are the part of the analyzed program itself, have a high Core Bound value, which usually indicates insufficient loading of functional units (e.g., due to data dependency) or restrictions imposed by some complex arithmetic operations, such as division, which is quite widely used in this application. You can also highlight the constructors of the \texttt{point} class, which are also appear to be Memory Bound functions.

It is worth mentioning the vectorization aspect of the code. In this program, the main operations are performed on integers, not on floating-point operations, and VTune does not allow assessing the quality of vectorization of such operations (for this purpose, a tool like Intel Advisor can be used). However, it is known that the vectorization in this program can be improved, and this issue will be addressed in the future.

Based on the performed analysis, the following conclusions can be drawn regarding the possibility of program optimization:
\begin{enumerate}
    \item MPI operations occupy only a small part of the program, and optimizing this aspect of the application will not bring a significant speedup.
    \item In addition to application functions, Linux system calls for working with paged memory (for example, \texttt{copy\_page\_rep}, \texttt{clear\_page\_c\_e}) take up significant time, and these calls show rather low performance. Thus, a possible optimization option is to modify the program to reduce the number of such calls, or to find more optimized implementations of these libraries.
    \item A number of program functions (like \texttt{mul\_inv}, \texttt{mul\_mod}) can leave CPU underutilized due to the use of operations with large delays (such as the division operation). Modern compilers can optimize division when using 32-bit numbers, but for 64-bit and 128-bit operands DIV, IDIV operations are generated, and the delay in this case can exceed 70 processor cycles, as well as the readiness for the next such operation. In some situations, it may be possible to change the code to use multiplication instead of division (if overflow problems are not expected), or to use approximate division implementations.
    \item Some functions (for example, \texttt{add\_to}) have a high Bad Speculation score, which is usually due to the Branch Prediction. In the case of the specified \texttt{add\_to} function, this can happen both due to the presence of a loop with a break condition or the presence of branching operators in the loop. Modifying this fragment can also help speed up the execution of the program.
    \item The program is not fully vectorized, and a more detailed study of this issue is also an option for further optimization.
    \item Some constructors of the point class are memory-bound functions -- they frequently access the memory of different arrays. For them, it is worth considering memory optimization (for example, improving data locality).
\end{enumerate}


After performance analysis, an optimized version of the program was developed. In the second version of the application, two modifications are introduced: 1) the constructor of the point class now uses bit storage to optimize memory management; 2) an approximate implementation of the division operation is used (the division accuracy is sufficient in this case, and the execution time of such an implementation of the operation is reduced). Other optimization options are expected to be tested in the future.

The average execution time of the optimized program decreased by 5 seconds, which is 4.3\% of the total execution time of the initial version.

The performance of the optimized version was analyzed using VTune and compared with the initial version. The analysis was carried out in the same conditions specified at the beginning of this section.

\begin{figure}[!h]
\includegraphics[width=1.0\textwidth]{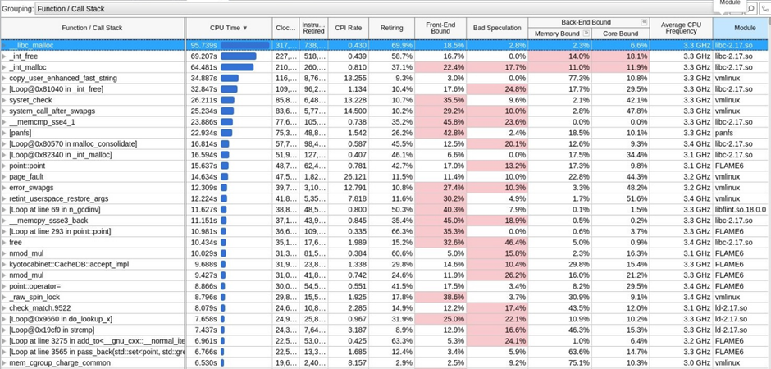}
\caption{The most computationally expensive program fragments in the optimized version of the application}
\label{fig:vtune-opt}
\end{figure}

Let us briefly describe the main conclusions. Although the speedup is not so notable, the efficiency metrics (useful utilization of CPU, as well as the average number of clock cycles per instruction) improved comparing to the initial program. Further, the list of the most computationally expensive program fragments (see Fig.~\ref{fig:vtune-opt}) changed significantly. Now it is mainly occupied by functions from the \texttt{glibc} library for allocating and freeing memory, such as \texttt{malloc} and \texttt{free}. Moreover, the fragments from the program itself (i.e., not external functions) first occur only at the 12th place in Fig.~\ref{fig:vtune-opt}, i.e., the top 11 most computationally expensive fragments are executed outside the body of the program itself. This indicates that for further optimization it is necessary to either use more optimized implementations of external functions (i.e. use other libraries) or reduce the number of external calls in the program, primarily related to memory allocation/freeing and working with memory pages. This will be done in the future.

\section{Optimization with the use of Flint library}
\label{FlintSect}

The use of the FLINT library \cite{Flint} in FIRE was already implemented in Ref.~\cite{Smirnov:2023yhb} through the library FUEL \cite{Mokrov:2023vva}. Initially, there was no support for rational functions with modular arithmetic since these were not used within the Beast mode. But after changing the reconstruction order, as was discussed in Section \ref{ReconOrderSect}, there arose an urgent need to efficiently perform such operations in the Beauty mode.

Since FLINT does not (yet) have an in-house implementation of modular rational functions, we wrote our own version on top of FLINT's routines for polynomials over prime fields. The FLINT routines include binary arithmetic operations (additions etc.) for such polynomials, polynomial GCD computations and exact division of polynomials when there is no remainder. We translated a similar implementation of modular rational functions from the Julia package Nemo \cite{Nemo} to C++. As an example, let us spell out the algorithm, translated from Nemo, for computing and simplifying the sum of two rational functions, $n_1 / d_1 + n_2 / d_2$, where $n_1, n_2, d_1, d_2$ are polynomials (over prime fields). We assume the two rational functions to have been simplified previously, which means $n_1$ and $d_1$ have no nontrivial polynomial GCD, and similarly for $n_2$ and $d_2$.
Various special cases, corresponding to {\tt if-else} statements in the actual code, are treated separately to ensure efficiency on a computer. We summarize the algorithm below.
\begin{itemize}
    \item Case 1: $d_1 = d_2$.
    
    The sum is equal to the new rational function $(n_1+n_2)/d_1$. If $d_1=1$, this is the returned result. Otherwise, we compute the polynomial GCD of $(n_1+n_2)$ and $d_1$, and divide the numerator and denominator by the GCD if the GCD is not equal to one, before returning the rational function.
    \item Case 2: $d_1 = 1$.

    The sum is equal to the new rational function $(n_2+d_2) / d_2$. Since $n_2$ and $d_2$ have no nontrivial polynomial GCD, $(n_2+d_2)$ and $d_2$ also have no nontrivial polynomial GCD, and the above rational function is returned.
    \item Case 3: $d_2 = 1$.

    The treatment is similar to case 2.
    \item Case 4: all other cases.

    We compute the polynomial GCD of $d_1$ and $d_2$ and let the result be $g$.
     \begin{itemize}
        \item Case 4(a): $g$ is equal to 1.
        
        The sum is $(n_1 d_2 + n_2 d_1) / (d_1 d_2)$. This result is returned, since it is easy to prove that the numerator and denominator cannot have a nontrivial GCD, under the starting assumption that $n_1/d_1$ and $n_2/d_2$ are previously simplified rational functions.
     \item Case 4(b): $g$ is a nontrivial polynomial.

        We let $q_1 = d_1 / g$ and $q_2 = d_2 / g$. The sum is $(n_1 q_2 + n_2 q_1)/ (q_1 q_2)$, and we further simplify this sum if there is a nontrivial GCD between the numerator and denominator to be canceled, before we return the result.
    \end{itemize}
\end{itemize}

Another performance improvement we have implemented is related to the communication between FIRE and the simplification library, in this case FLINT. All communications are via strings: FIRE sends strings of rational functions expressions to FLINT, while FLINT (with the help of our parser code) reads the strings and sends back the simplified rational functions, again as strings. Since simplified rational functions will be used as sub-expressions in subsequent calculations, long expressions may be re-sent to the simplification library, causing a re-evaluation before further calculations. This overhead is avoided if the expression is stored in memory in an internal format native to the simplification library. To accomplish this without a dramatic rewrite of the FIRE database, we retained the use of strings as the bidirectional communication format, while adding special syntax to the strings, support4ed by our parser, to refer to expressions that are stored in memory. Specifically, the expressions are stored in a hash map that maps its numerical label (1, 2, 3, etc.) to the actual expression in a native (i.e.~non-string) format. Imagine two previously simplified rational functions, say, $(1-x)/(1+x^2)$ and $x/(1-x^2)$, have to be added in a subsequent calculation, previous versions of FIRE will send the string $(1-x)/(1+x^2) + x/(1-x^2)$ to the simplifier, while the current version of FIRE can optionally send a string like \{1\}+\{3\} to mean adding the 1st and 3rd expressions stored in memory.\footnote{The actual implementation requires some memory management such as periodic cleaning of unneeded stored expressions to prevent memory usage from growing out of control.} This has resulted in dramatic performance improvements. In our four-variable benchmark example described in the last paragraph of Section 
\ref{ReconOrderSect}, the measurement of times needed for rational function reconstruction of one-to-four variables in the Beauty mode is shown in Table \ref{FlintTable}.

\begin{table}[]
    \centering
    \begin{tabular}{|l|c|c|c|}
    \hline
    vars/reconstruction & Old Way & New Way & With storing \\
    \hline
        $d$/Thiele          & 0.06 &  0.06  &  0.06 \\
        $(d,w)$/Balanced Newton & 2.49 &  1.77  &  0.6  \\
        $(d,w,v)$/Balanced Newton & 85 &  39.6  &  6.28 \\
        $(d,w,v,u)$/Balanced Newton & 5213 &  2346  &  402 \\
   \hline
    \end{tabular}
    \caption{Runtimes (in seconds) for rational reconstructions. 
    The `Old Way' corresponds to the initial implementation of  communications with FLINT in FUEL --- non-modular functions were called first and then projection of coefficients to the proper modular field were taken. The `New Way' is the current approach. The use of storing speeds it up things even more.}
    \label{FlintTable}
\end{table}

\section{Conclusions}

In this paper, we devised a number of various techniques for improving performance of Feynman integral reduction on supercomputers. Cumulatively, with all methods applied, we estimate the resource economy more that twice compared with the version without their implementation. In IBP reductions started from scratch, initial optimization can be achieved by a proper of choice of the order of seeding variables to be restored, as was already discussed in Ref.\ \cite{Belitsky:2023qho}: the rule of thumb being to starting from a variable which requires less Thiele/balanced Newton BBSs and gradually proceeding to the ones which need more. This is highly problem-specific indeed. In this paper, all runs for our benchmark example were done for the optimal order of variable reconstruction $d \to w \to v \to u$ from the get-go. Future improvements include developing a hybrid technique of balancing and Zippel \cite{Zippel79,Zippel90} reconstruction methods, which are particularly suited for sparse polynomials. The achieved performance gains open up possibilities to apply {\tt FIRE} on supercomputers to a number of cutting-edge physical problems including minimal and four-leg form factors of the stress-tensor multiplet, and five- and six-leg scattering amplitudes on the Coulomb branch of the maximally supersymmetric Yang-Mills theory at two loops, just to name a few.

\begin{acknowledgments}
A.B.\ is grateful to 
Ayush Saurabh and Gil Speyer for the initial introduction to ASU's {\tt Sol} supercomputer and Vladimir A.~Smirnov for useful discussions.
The work of A.B.\ was supported by the U.S.\ National Science Foundation under the grant No.\ PHY-2207138. 
The work of A.S., in part of developing improved reconstruction algorithms, was supported by the Ministry of Education and Science of the Russian Federation as part of the program of the Moscow Center for Fundamental and Applied Mathematics under Agreement No.\ 075-15-2022-284. The work of A.S. and V.V., in part of optimizing supercomputer algorithms, was supported by the Russian Science Foundation under Agreement No.\ 21-71-30003.

The research was carried out using the equipment of the shared research facilities of HPC computing resources at Lomonosov Moscow State University \cite{Lomonosov}. 
Performance benchmarking was also performed at the Polus supercomputer \cite{Polus}.
\end{acknowledgments}

\end{document}